\documentclass[conference,12pt]{IEEEtran}
\usepackage[T1]{fontenc}
\usepackage{txfonts}

\newcommand{\ket}[1]{|#1\rangle}

\author{\IEEEauthorblockN{Ryutaroh Matsumoto}
\IEEEauthorblockA{Department of Information and Communication Engineering\\
Nagoya University, Japan}}
\title{Exploring Quantum Supremacy in Access Structures of Secret Sharing
by Coding Theory}

\begin{document}
\maketitle
\begin{abstract}
  We consider secret sharing schemes with a classical secret and
  quantum shares. One example of such schemes
  was recently reported whose access structure cannot be realized by
  any secret sharing schemes with classical shares.
  In this paper, we report further quantum secret sharing schemes
  whose access structures cannot be realized by any 
  classical secret sharing schemes.
\end{abstract}

\section{Introduction}
Secret sharing \cite{shamir79}
is a cryptographic scheme to encode a secret
into multiple pieces of information (called shares) so that
only qualified sets of shares can reconstruct the original
secret.
Secret sharing has become even more important
as its application to the cloud storage is spreading \cite{attasena2017}.
The security criterion of secret sharing is usually information
theoretic one and thus cannot be broken even by quantum computers
\cite{stinson06}.

Quantum supremacy \cite{preskill12} is
the potential ability of quantum computing devices to solve problems that classical computers practically cannot.
Discovery of new quantum supremacy is important in research of quantum
information processing.
Since majority of secret sharing schemes are secure against both classical and quantum computers,
quantum supremacy cannot be found in that respect.
On the other hand, the author recently reported new quantum
supremacy in the access structure of secret sharing \cite{matsumoto17impossible}.
An access structure of a secret sharing schemes
is a set of qualified share sets and forbidden share sets,
where a share set is said to be forbidden (resp.\ qualified)
if the set has no information about the secret (resp.\
can reconstruct the secret) \cite{ogawa05}.

Specifically, when we use the famous $[[5,1,3]]$ binary quantum
stabilizer error-correcting code to encode a 1-bit classical secret
into five quantum shares, its access structure cannot be realized by
any secret sharing schemes with classical shares.
However, it was not clarified whether or not there exists another
secret sharing schemes with quantum shares whose access structures
cannot be realized by classical shares.
In this paper, we use different necessary conditions on
the existence of access structures realized by
secret sharing schemes with classical shares, and report 9 new quantum secret sharing
schemes whose access structures cannot be realized by 
secret sharing schemes with classical shares.

\section{Quantum Error-Correcting Codes and Secret Sharing}
Quantum error-correcting codes have been used for constructing secret sharing schemes
for quantum secrets \cite{cleve99,marin13,matsumoto17uni}.
Since classical information can be regarded as a special case of
quantum information \cite{chuangnielsen}, it is easy to
construct a secret sharing scheme for a classical secret
from a quantum error-correcting code.
Suppose that we have a $k$-bit string $\vec{s}$ as a classical secret and
we want to encode $\vec{s}$ into $n$ shares.
For this goal, we select a binary $[[n,k,d]]$ quantum error-correcting code $Q$,
where $[[n,k,d]]$ means that the code encodes $k$ qubits into $n$ qubits and
has the minimum distance $d$.
We prepare a $k$-qubit quantum state $\ket{\vec{s}}$ and
encode $\ket{\vec{s}}$ into $n$ qubits $\ket{\vec{x}}$ by $Q$.
Then each qubit in the quantum codeword $\ket{\vec{x}}$ is distributed to
each of $n$ participants.

We say that a secret sharing scheme has
$t$-privacy if any set of $t$ shares has absolutely
no information about the secret,
and has $r$-reconstruction if any set of $r$ shares
uniquely reconstruct the secret \cite{cascudo18}.
For simplicity, $r$ is assumed to be smallest possible and
$t$ to be largest possible.
For a secret sharing scheme to be useful,
we must know $r$ and $t$.
We will relate $r$ and $t$ in order to demonstrate
the quantum supremacy.

\section{Quantum Supremacy in Access Structures}
Suppose that one has $n-d+1$ or more shares.
Then the number of missing shares is $d-1$ or less.
By setting the quantum state of missing shares to any state (e.g., the completely mixed state)
and treating them as erasures, the quantum erasure correction procedure
reconstructs the $n$ shares $\ket{\vec{x}}$ from available shares
\cite{matsumoto17impossible}, and the secret $\vec{s}$
can be reconstructed from $\ket{\vec{x}}$.
This means that $r \leq n-d+1$.

On the other hand,
when we have a secret sharing scheme with a classical secret and quantum shares and
a set of shares can reconstruct the secret,
then the complementary set of shares has absolutely no information about the secret \cite{ogawa05}.
This implies that $t \geq d-1$.

The difference $r-t$ is called the threshold gap.
When we construct a secret sharing scheme from a binary $[[n,k,d]]$ quantum error-correcting codes,
we have
\begin{equation}
  r-t \leq n+2 - 2d. \label{eq1}
\end{equation}

On the other hand,
when we have a secret sharing scheme in which
each classical share has $\log_2 q$ bits and the classical secret has
$k \log_2 q$ bits, we must have \cite{bogdanov16}
\begin{equation}
  r-t \geq \frac{r+1}{q}. \label{eq2}
\end{equation}

A secret sharing scheme with classical shares is said to be
\emph{linear} if the reconstruction from shares to secrets is a linear map \cite{chen07}.
Most of studied secret sharing schemes with classical shares
are linear, as they enable efficient encoding and reconstruction by
linear algebraic algorithms.
When a scheme is linear, we must have \cite{cascudo18}
\begin{eqnarray}
  && r-t \geq \frac{q^m-1}{q^{m+1}-1}(n+2) + \frac{q^{m+1}-q^m}{q^{m+1}-1}(k - 2m)\nonumber\\
  && \mbox{ (for all $0\leq m \leq k-1$)}.\label{eq3}
\end{eqnarray}

We consider the case that each share is one bit or one qubit,
and search for an access structure that can be realized by quantum shares but
cannot be realized by classical shares.
If we have a binary $[[n,k,d]]$ quantum code and we also have
\begin{equation}
  n+2-2d < \frac{n+2-d}{2}, \label{eq4}
  \end{equation}
then by Eqs.\ (\ref{eq1}) and (\ref{eq2}) the 
binary $[[n,k,d]]$ quantum code realizes an access structure
that cannot be realized by secret sharing schemes with classical 1-bit shares,
thus it exhibits quantum supremacy in the access structure.

In addition, if we have a binary $[[n,k,d]]$ quantum code and we also have
\begin{eqnarray}
&&  n+2-d < \frac{q^m-1}{q^{m+1}-1}(n+2) + \frac{q^{m+1}-q^m}{q^{m+1}-1}(k - 2m) \nonumber\\
&&  \mbox{ (for some $0\leq m \leq k-1$)}, \label{eq5}
\end{eqnarray}
then by Eqs.\ (\ref{eq1}) and (\ref{eq3}) the 
binary $[[n,k,d]]$ quantum code realizes an access structure
that cannot be realized by \emph{linear} secret sharing schemes with classical 1-bit shares,
thus it also exhibits quantum supremacy in the access structure.

Grassl \cite{codetable} maintains the table of best binary quantum error-correcting codes.
We searched for codes with properties (\ref{eq4}) or (\ref{eq5}), and
found the codes in Table \ref{tab1}.

\begin{table}
  \caption{Parameters of binary $[[n,k,d]]$ quantum error-correcting codes
    that exhibit quantum supremacy in the access structures of associated secret sharing
    schemes}\label{tab1}
\centering  \begin{tabular}{cccll}\hline
    $n$&$k$&$d$&Eq. (\ref{eq4})&Eq. (\ref{eq5})\\\hline
    6&1&3&true&false\\
    11&1&5&true&false\\
    12&1&5&true&false\\
    17&1&7&true&false\\
    18&1&7&true&false\\
    27&3&9&false&true with $m=2$\\
    28&3&9&false&true with $m=2$\\
    29&1&11&true&false\\
    30&1&11&true&false\\\hline
  \end{tabular}
\end{table}

\section{Conclusion}
As a continuation of the author's recent paper \cite{matsumoto17impossible},
we searched quantum error-correcting codes that give secret sharing schemes
whose access structures cannot be realized by classical information processing.
We reported 9 new codes having access structures impossible by classical information
processing in Table \ref{tab1}.
However, it remains unknown whether or not there exist infinitely many
quantum error-correcting codes having access structures impossible by classical information processing.
It is a further research agenda.

\section*{Acknowledgment}
This research is partly supported by the JSPS Grant No.\ 26289116.



\end{document}